\documentclass[pre,showpacs,twocolumn,floatfix,amsmath,amsfonts]{revtex4}

\newcommand{\g}{\gamma}

\newcommand{\iu}{\mathrm{i}}

\def\om{\omega_k}

\usepackage{epsfig}
\usepackage{graphicx}
\usepackage{amsmath}
\usepackage{amssymb}
\usepackage{dcolumn}

\usepackage{dsfont, srcltx}

\begin{document}

\title{Resonant interaction of $\phi^4$ kink with spatially periodic $\mathcal{PT}$-symmetric perturbation}

\author{Danial~Saadatmand$^{1}$}
\email{saadatmand.d@gmail.com}

\author{Denis~I.~Borisov$^{2,3,4}$}
\email{borisovdi@yandex.ru}

\author{Panayotis~G.~Kevrekidis$^{5}$}
\email{kevrekid@math.umass.edu }

\author{Kun~Zhou$^{6}$}
\email{kzhou@ntu.edu.sg}

\author{Sergey~V.~Dmitriev$^{7,8}$}
\email{dmitriev.sergey.v@gmail.com}

\affiliation{$^1$Department of Physics, University of Sistan and Baluchestan, Zahedan, Iran
\\
$^2$Institute of Mathematics, Ufa Scientific Center, Russian Academy of Sciences, Chernyshevsky 112, 450008 Ufa, Russia
\\
$^3$Bashkir State Pedagogical University, Oktyabrskoy Revolutsii 3a, 450000 Ufa, Russia
\\
$^4$ University of Hradec Kr\'alov\'e, Rokitansk\'eho 62, 500 03 Hradec Kr\'alov\'e, Czech Republic
\\
$^5$Department of Mathematics and Statistics, University of Massachusetts, Amherst, MA 01003 USA
\\
$^6$School of Mechanical and Aerospace Engineering, Nanyang Technological University, 50 Nanyang Avenue, Singapore 639798, Singapore
\\
$^7$Institute for Metals Superplasticity Problems RAS, Khalturin 39, 450001 Ufa, Russia
\\
$^8$Saint Petersburg State Polytechnical University, Politekhnicheskaya 29, 195251 St. Petersburg, Russia
 }

\begin{abstract}
The resonant interaction of the $\phi^4$ kink with a periodic $\mathcal{PT}$-symmetric perturbation is observed in the frame of the continuum model and with the help of a two degree of freedom collective variable model derived in PRA {\bf 89}, 010102(R). When the kink interacts with the perturbation, the kink's internal mode is excited with the amplitude varying in time quasiperiodically. The maximal value of the amplitude was found to grow when the kink velocity is such that it travels one period of perturbation in nearly one period of the kink's internal mode. It is also found that the kink's translational and vibrational modes are coupled in a way that an increase in the kink's internal mode amplitude results in a decrease in kink velocity. The results obtained with the collective variable method are in a good qualitative agreement with the numerical simulations for the continuum model. The results of the present study suggest that kink dynamics in open systems with balanced gain and loss can have new features in comparison with the case of conservative systems.
\end{abstract}
 \maketitle

\section {Introduction}

In physics, resonance is a ubiquitous phenomenon. At resonant frequencies, small periodic driving forces have the ability to produce large amplitude oscillations because the system gradually accumulates vibrational energy, given a
pathway of exchange between intrinsic vibration and external drive.
Resonance effects have been investigated for wave scattering
processes in different branches of physics \cite{Fano}.

A widely studied example of resonance effects has been explored in the
realm of (non-integrable)
field theoretic models such as nonlinear Klein-Gordon partial differential
equations
in the context of their soliton dynamics. Such effects are often related to the feature that the kinks (topological excitations) in non-integrable Klein-Gordon models can support vibrational internal modes  (IM)~\cite{KivsharIM,pgkIM,Quintero,Belova} and that the energy exchange between the kink's translational and vibrational modes is possible~\cite{C1,C2,C3,C4,C5,pgkin,R5,Gani1,Gani2}. For the $\phi^4$, $\phi^6$, and $\phi^8$ equations the multibounce resonances in the kink-antikink collisions have been observed and have been
attributed to the energy exchange between kink translational and vibrational modes \cite{C1,C2,C3,C4,C5,R5,Gani1,Gani2}. Similar effects are also possible in the $\phi^6$ model in the absence of the kink's IM due to the existence of bound states in the spectrum of small oscillations about a combined kink-antikink configuration \cite{R6}; interestingly,
such features were also present in the case of the parametrically modified
sine-Gordon equation despite the potential absence of IMs in that
case as well~\cite{Quintero}. In the context of IM considerations, however,
it is also timely and relevant to mention that recent works have
disputed the validity and quantitative accuracy of such collective
coordinate models~\cite{Weigel}.
When more than two Klein-Gordon kinks collide, the multibounce resonances and fractal soliton scattering are possible due to the radiationless energy exchange between the kinks in near-separatrix collisions \cite{S1,S2,S3}. Vibrational impurity modes in the $\phi^4$ and sine-Gordon equations can also result in multibounce resonances in the kink-impurity interactions \cite{R1,R2,R3}. Kink dynamics in spatially modulated $\phi^4$ model, $\phi_{tt}-\phi_{xx}+[1+\epsilon\sin(\kappa x)](\phi^3-\phi)=0$, have been analysed revealing the four types of kink behavior depending on its velocity: (i) small-amplitude wave radiation at high velocities, (ii) strong resonant beating in the kink velocity at somewhat smaller velocities, (iii) propagation of the kink with almost periodic velocity oscillations at even smaller velocities, and (iv) trapping at very low velocities \cite{SpaceModulation}. Note that in the above mentioned works the energy conserving systems have been analysed.

Very often the resonant soliton dynamics are observed in the Klein-Gordon models perturbed by ac external driving combined with damping, i.e., in the open systems with energy gain and loss. The most typical example is the $\phi^4$ equation  $\phi_{tt} - \phi_{xx} + \phi^3 - \phi=-\gamma \phi_t + \epsilon\sin(\omega t+\delta_0)$, with the damping constant $\gamma$ and harmonic ac force of small amplitude $\epsilon$, frequency $\omega$, and initial phase $\delta_0$. In the
context of this model, in the damped case, the strongest resonance has been found at half the frequency of the kink's IM \cite{acdriven}. The sine-Gordon model with spatiotemporal perturbations, $\phi_{tt} - \phi_{xx} + \sin \phi=-\gamma \phi_t + F(x,t)$ either with some spatially inhomogeneous forces $F(x)$ or with spatiotemporal inhomogeneous forces $F(x,t)$, also manifests interesting kink dynamics~\cite{sptemp}. The influence of the IM on the soliton ratchet mobility has been investigated for the driven and damped double sine-Gordon equation $\phi_{tt} - \phi_{xx} =-dU/d\phi - \gamma \phi_t + f(t)$, with $U=(1-\cos\phi)+(\lambda/2)(1-\cos 2\phi)$, $\lambda$ is a parameter, and $f(t)= \epsilon_1 \sin(\omega t +\delta_1) + \epsilon_2 \sin(2\omega t +\delta_2)$ \cite{dsge}. In this case,
the kink's IM constantly radiates its energy. The compensation of the radiation losses of the kink's IM by the resonant driving of the kink has been demonstrated for the model $(\phi_{tt} - \phi_{xx})/2 + \gamma \phi_t - [1+\epsilon\cos(\omega t)]\phi + \phi^3 = 0$ \cite{OB}. The ratchet effect of a sine-Gordon kink has been investigated in the absence of any external force while the symmetry of the field potential at every time instant is maintained. This was done for the model $\phi_{tt} - \phi_{xx} + \gamma \phi_t + U^\prime (\phi, t) = 0$ with the time-dependent potential $U(\phi, t) = 1 - \cos[\phi + \theta \eta(t)]$, where $\theta$ is the amplitude of the periodic function $\eta(t)$ having zero time average \cite{niu}.

Recently, Bender and co-authors have offered a class of open systems described by non-Hermitian Hamiltonians possessing real spectra under the parity-time ($\mathcal{PT}$) symmetry condition, where parity-time means spatial reflection and time reversal \cite{Bender1,Bender2}. Such open physical systems with
balanced gain and loss have now been realized experimentally in optics
\cite{Ruter,Peng2014,peng2014b,RevPT,Konotop}, electronic circuits
\cite{Schindler1,Schindler2,Factor}, and mechanical systems \cite{Bender3}.
This intense line of research activity has recently been summarized in
two comprehensive reviews~\cite{RevPT,Konotop}.

The Klein-Gordon field with a $\mathcal{PT}$-symmetric localized perturbation describing a defect with balanced gain and (viscous) loss has been recently introduced by one of the authors~\cite{KevrekidisRevA}. Scattering of the kink from a spatially localized $\mathcal{PT}$-symmetric defect was recently investigated in the $\phi^4$ model \cite{Danial1,Danial2}. The effect of the kink's IM is also discussed in these papers. It was demonstrated that if a kink hits the defect from the gain side, a noticeable IM is excited, while for the kink coming from the opposite direction the mode excitation is much weaker. The interaction of the moving kinks and breathers with the spatially localized $\mathcal{PT}$-symmetric perturbation was also investigated in the realm of the sine-Gordon field \cite{Danial}. Several new soliton-defect interaction scenarios were observed such as the kink passing/being trapped depending on whether the kink comes from the gain or loss side of the impurity, merger of the kink-antikink pair into a breather, and splitting of the breather into a kink-antikink pair. In \cite{Danial1}, it has been shown that if the kink interacts with a $\mathcal{PT}$-symmetric defect, one may expect a significant IM excitation. An intriguing question
that we intend to explore herein is what happens if a moving $\phi^4$ kink interacts with a periodic $\mathcal{PT}$-symmetric perturbation~? In the following we will see that at certain kink velocities there exists a resonance effect in which case the maximal amplitude of the kink's IM increases considerably.

The outline of the paper is as follows. In Sec.~\ref{Sec:II} the simulation setup is described. We report on the numerical results presenting the interaction of the kinks with the periodic $\mathcal{PT}$-symmetric perturbation in Sec.~\ref{Sec:III}. A collective variable approach is developed and used in Sec.~\ref{Sec:IV}. Our conclusions are given in Sec.~\ref{Sec:V}.

\section {Simulation setup} \label{Sec:II}

In this section we introduce the spatially localized periodic $\mathcal{PT}$-symmetric
inhomogeneity into the $\phi^4$ field, present the approximate solution to the unperturbed $\phi^4$ model describing a moving kink bearing the kink's IM, and derive the resonance condition for the kink's IM interacting with the periodic inhomogeneity.

\subsection {The model} \label{Sec:IIA}

We study the modified $\phi^4$ equation \cite{KevrekidisRevA}
\begin{equation}\label{phi4}
\frac{\partial^2\phi}{\partial t^2} - \frac{\partial^2\phi}{\partial x^2} -2\phi(1-\phi^2) = \gamma(x)\frac{\partial\phi}{\partial t},
\end{equation}
where $\phi(x,t)$ is the unknown scalar field and
\begin{equation}\label{Phi4_Perturbation}
\gamma(x)=\epsilon\sin(\beta x),
\end{equation}
where $\epsilon$ is the amplitude and $\beta$ defines the spatial period of the perturbation term,
\begin{equation}\label{lambda}
\lambda=\frac{2\pi}{\beta}.
\end{equation}
Note that $\gamma(-x)=-\gamma(x)$. Under this condition equation~(\ref{phi4}) is $\mathcal{PT}$-symmetric, i.e., it preserves its form under the change $x\mapsto -x$ and $t\mapsto-t$. From the physical standpoint Eq.~(\ref{phi4}) describes an open system with (periodically) balanced gain and loss. In the loss (gain) regions $\gamma(x)<0\,(>0)$.

For $\epsilon=0$, Eq.~(\ref{phi4}) is the non-integrable $\phi^4$ equation with the following exact moving kink solution
\begin{equation}\label{Kink}
   \phi_K(x,t)=\pm\tanh(\delta_k (x-x_0-V_{k}t)),
\end{equation}
where
\begin{equation}\label{delta}
\delta_{k} =\frac{1}{\sqrt{1 - V_{k}^2}}\,,
\end{equation}
$V_k$ is the kink velocity and $x_0$ is the kink initial position.

\begin{figure}
\includegraphics[width=9.5cm]{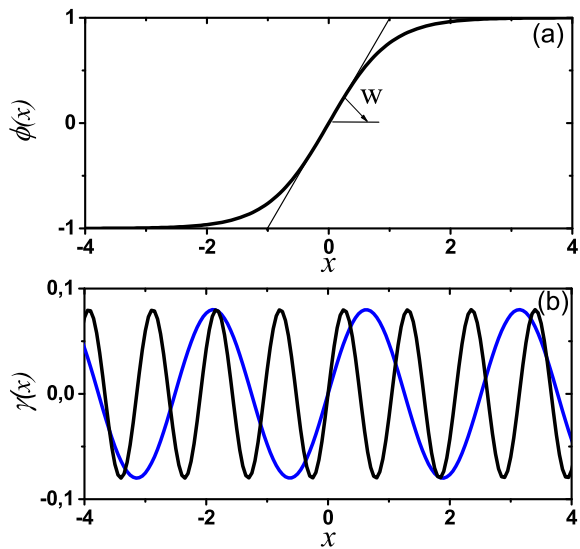}
\caption{Setting the stage: the coherent structure and the gain-loss
  ``terrain'' in
  which it moves. (a) Static kink solution Eq. (\ref{Kink}) to the non-perturbed $\phi^4$ equation ($\epsilon=0$) for $x_0=0$ and velocity $V_k=0$. The definition of the inverse kink width, $W$, is shown. (b) The perturbation term Eq.~(\ref{Phi4_Perturbation}) for $\epsilon=0.08$ and $\beta=2.5$ (blue line) and $\beta=6.0$ (black line).} \label{fig1}
\end{figure}

In Fig.~\ref{fig1}(a) the kink solution Eq.~(\ref{Kink}) is shown for the initial position $x_0=0$ and velocity $V_k=0$. In (b) the function $\gamma(x)$ given by Eq.~(\ref{Phi4_Perturbation}) is plotted for the amplitude $\epsilon=0.08$ and two different periods corresponding to $\beta=2.5$ (blue line) and $\beta=6.0$ (black line).

\subsection {Kink's internal mode} \label{Sec:IIB}

It is well-known that the $\phi^4$ kink possesses a long-lived IM
in the linearization around the kink~\cite{sugiyama}. In the
Appendix we demonstrate that the kink bearing an IM of amplitude $A\ll~1$ can be approximately presented in the form
\begin{equation}\label{Ansatz}
\begin{aligned}
&\phi_A(x,t)=\pm \tanh(\delta_k (x-x_0-V_k t))
 \\
& + A\cos\bigg\{\sqrt {3}\delta_k [t-V_k(x-x_0)]\bigg\}
\frac{\sinh(\delta_k (x-x_0-V_k t))}{\cosh^2 (\delta_k (x-x_0-V_k t))}
 \\
&\pm A^2\cos\bigg\{2\sqrt{3}\delta_k [t-V_k(x-x_0)]\bigg\}
\Phi_2(\delta_k (x-x_0-V_k t))
\\
&+O(A^3),
\end{aligned}
\end{equation}
where $\Phi_2$ is a solution to Eq.~(\ref{d10}).

To reveal the physical meaning of the IM, we calculate $\partial\phi_A/\partial x$ at the kink center, i.e., at $x=0$. The result reads
\begin{equation}\label{Kinkomega2}
   W:=\frac{\partial\phi_A}{\partial x} \Big\vert_{x=0} =\delta_k\bigg(1+A\cos\bigg\{\frac{\sqrt 3}{\delta_k} [t-V_k(x-x_0)]\bigg\}\bigg),
\end{equation}
where the small term proportional to $A^2$ was dropped.

One can regard $W$ as characterizing the inverse kink width [see Fig.~\ref{fig1}(a)], which in the absence of the IM (for $A=0$) is constant in time,
\begin{equation}\label{Kinkwidth}
   W_0=\delta_k,
\end{equation}
and increases with the kink velocity (the kink width characterized by
$1/W$ decreases with kink velocity). When the IM is excited, the inverse kink width oscillates near $W_0$
with the amplitude
\begin{equation}\label{Kinkampl}
   a=A\delta_k,
\end{equation}
and frequency
\begin{equation}\label{Kinkomega}
   \omega=\frac{\sqrt{3}}{\delta_k},
\end{equation}
which depend on the kink velocity, $V_k$.

\subsection {Numerical scheme} \label{Sec:IID}

To integrate Eq.~(\ref{phi4}) numerically, the mesh $x_n~=~nh$ is introduced, where $h$ is the lattice spacing, $n~=~0,\pm 1, \pm 2$,... and the following discrete version of the equation is proposed
\begin{eqnarray}\label{phi4discrete}
&&\frac{d^2\phi_{n}}{d t^2} - \frac{1}{h^2}(\phi_{n-1}-2\phi_{n}+\phi_{n+1}) \nonumber \\
&&+\frac{1}{12h^2}(\phi_{n-2}-4\phi_{n-1}+6\phi_{n}-4\phi_{n+1}+\phi_{n+2}) \nonumber \\
&&-2\phi_n(1-\phi_n^2)- \gamma_n\frac{d\phi_n}{d t}=0,
\end{eqnarray}
in which $\phi_n = \phi(nh,t)$ and $\gamma_n = \gamma(nh)$.
This discretization method is fourth order accurate in space.
The equations of motion Eq.~(\ref{phi4discrete}) were integrated with respect to the temporal variable using an explicit scheme with the accuracy of $O(\tau^4)$ and the time step $\tau$. The simulations were conducted for $h =0.05$ and $\tau =0.005$.

In the present study most of the simulations are carried out for the perturbation amplitude $\epsilon=0.08$. Only in Sec.~\ref{Sec:IIIC} the effect of the perturbation amplitude $\epsilon$ is considered.

\subsection {Kink's kinetic energy} \label{Sec:IIBB}

For the further discussion it is important to analyse kinetic energy of the kink moving with velocity $V_k$ and bearing internal vibrational mode. For this we calculate time evolution of the kink's kinetic energy
\begin{equation}\label{KinkKinEn}
   K=\frac{1}{2}\int_{-\infty}^{\infty}\left( \frac{\partial \phi}{\partial t} \right)^2 dx.
\end{equation}

For setting initial conditions Eq.~(\ref{Ansatz}) without the term proportional to $A^2$ is used. Here we analyze kink's kinetic energy for the unperturbed $\phi^4$ model ($\epsilon=0$). 

\begin{figure}
\includegraphics[width=10cm]{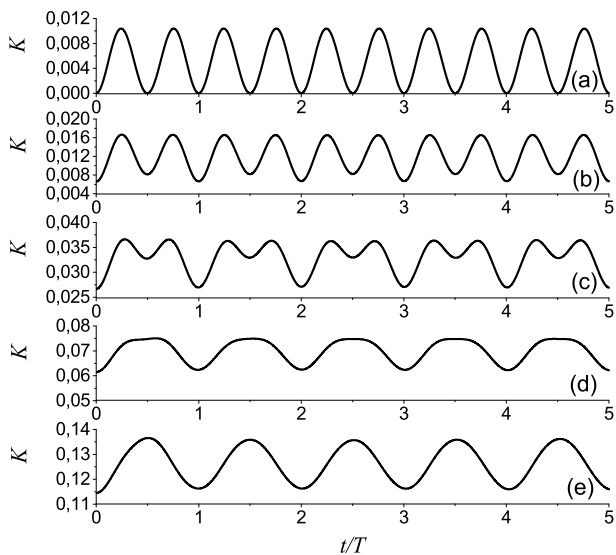}
\caption{Kinetic energy of the kink bearing kink's internal mode of amplitude $A=0.1$ as the function of time for (a)~$V_k=0$, (b)~$V_k=0.1$, (c)~$V_k=0.2$, (d)~$V_k=0.3$, and (e)~$V_k=0.4$. Time is normalized by the kink's internal mode period $T$. Case of unperturbed $\phi^4$ model ($\epsilon=0$).} \label{figK_t}
\end{figure}

In Fig.~\ref{figK_t} kinetic energy of the kink bearing IM with the amplitude $A=0.1$ is shown as the function of time for different kink velocities: (a)~$V_k=0$, (b)~$V_k=0.1$, (c)~$V_k=0.2$, (d)~$V_k=0.3$, and (e)~$V_k=0.4$. Time is normalized by the kink's internal mode period $T=2\pi/\omega=2\pi \delta_k/\sqrt 3$. It can be seen that in (a), for standing kink, $K(t)$ is a periodic function with the period equal to $T/2$, i.e., to the half of the kink's IM period. However for $V_k>0$ period of $K(t)$ is equal to $T$. Importantly, there is a critical value of kink velocity $V_k$ below which $K(t)$ has two maxima within $T$, while above this critical velocity it has only one maximum. For the chosen kink's IM amplitude of $A=0.1$ critical velocity is equal to 0.29, so that the $K(t)$ dependence shown in (d) is for the velocity slightly above the critical value. 

The critical velocity increases with the kink's IM amplitude $A$. Thus, for $A=0.15$ it is equal to 0.32, for $A=0.2$ it is 0.34 and for $A=0.25$ it is 0.35. The resonances observed in the present work happen for the above-critical kink velocities, when $K(t)$ has one maximum within $T$, and the reason for that is explained in Sec.~\ref{Sec:IIC}.

\subsection {Resonance condition} \label{Sec:IIC}

A perturbation of the ``dashpot'' type considered in this work is sensitive to $\partial \phi /\partial t$ and thus, the effect of the perturbation should be maximal when the kinetic energy of the kink is maximal. In the absence of perturbation, a kink bearing an excited IM demonstrates two maxima of its
kinetic energy within one IM period if it moves with a small velocity. For relatively large kink velocity there is only one maximum of the kinetic energy within one period of IM. We now proceed to derive the resonance condition for the fast kinks.

The kink moving with the velocity $V_k$ travels one period of the sinusoidal perturbation in time $t=\lambda/V_k=2\pi/(\beta V_k)$.
The (unperturbed) motion of the kink's center is described by the formula $x=x_0+V_k t$. For such $x$, the cosine in the second term in (\ref{Ansatz}) can be rewritten as
\begin{align*}
\cos (\sqrt{3}\delta_k [t-V_k(x-x_0)])=& A\cos (\sqrt{3}\delta_k [t-V_k^2 t])\\
=& A\cos \bigg(\frac{\sqrt{3}}{\delta_k} t\bigg)
\end{align*}
Hence, the period of the kink's IM is $T=
2\pi \delta_k/\sqrt 3$. Letting $t=T$ we obtain the resonance condition $\beta=\sqrt 3/(V_k \delta_k)$, from which, taking into account Eq.~(\ref{delta}),
\begin{equation}\label{Resonance1}
   \beta=\frac{\sqrt{3(1-V_k^2})}{V_k},
\end{equation}
or
\begin{equation}\label{Resonance2}
   V_k=\sqrt{\frac{3}{3+\beta^2}}.
\end{equation}
I.e., under this condition, the kink's traversing of the gain-loss
``terrain'' occurs on the same time scale as its intrinsic IM oscillation,
hence enabling the energy exchange between the two processes.

For slow kinks bearing IM there are two maxima of kinetic energy within one IM period and the resonance condition should be derived from $2t=T$. In this case Eq.~(\ref{Resonance1}) transforms to $\beta=2\sqrt 3 \delta_k/V_k$. But for slow kinks ($V_k \ll 1$) from these considerations, one has $\beta \approx 2\sqrt 3/V_k \gg 1$. The resonance condition can be satisfied only in the case of
the period of the perturbative function being much less than the kink width. Obviously, the resonance effect will be weak in this case because the kink will spread over the region with many alternating positive and negative damping regions.

\section {Numerical Results} \label{Sec:III}

In this Section, we firstly consider the unperturbed $\phi^4$ equation ($\epsilon=0$) and verify the accuracy of the ansatz of Eq.~(\ref{Ansatz}),
where appropriate, omitting $A^2$ term for simplicity. To do so, we excite kink's IM with the ansatz parameters $V_k$ and $A$ and find numerically actual kink velocity $\overline{V}_k$ and kink's IM amplitude $\overline{a}$.
Then, the kink dynamics is studied numerically in the full model of
Eq.~(\ref{phi4}).
Finally, the effect of the perturbation amplitude $\epsilon$ is analyzed.

\subsection {Kink's internal mode parameters in the unperturbed $\phi^4$ model} \label{Sec:IIIA}

\begin{figure}
\includegraphics[width=9.5cm]{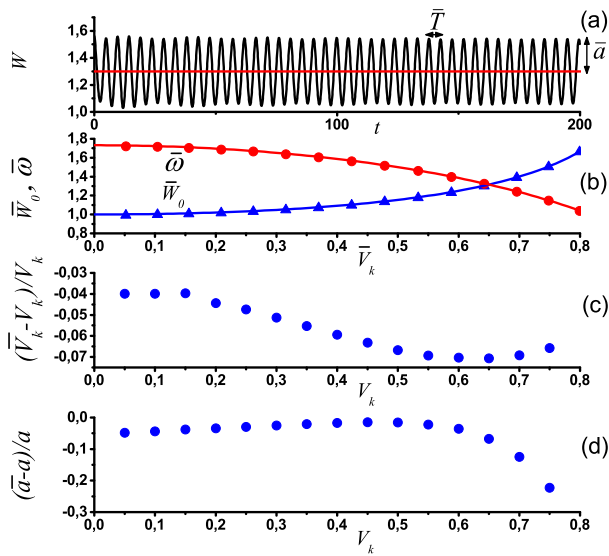}
\caption{Characteristics of the kink's internal mode in the unperturbed $\phi^4$ equation. (a) The kink's inverse width as a function of time for the following parameters of the ansatz in Eq.~(\ref{Ansatz}): $V_k=0.6$, $A=0.25$.
  (b) Variation of the values of the average inverse width and of the
  IM frequency
as a function of the speed $\overline{V}_k$; (c) deviation of the kink
speed numerically identified from the corresponding theoretically
prescribed one as a function of initial speed. Finally, (d) shows the
deviation of the IM amplitude from the corresponding prescribed value
thereof. See also the discussion in the text about these deviations.} \label{fig3}
\end{figure}

We use the ansatz of Eq.~(\ref{Ansatz}) (not taking into account the $A^2$ term) with the parameters $V_k$ and $A$ for setting initial conditions of
a moving kink bearing an excited  IM. Then we integrate numerically the unperturbed $\phi^4$ equation ($\epsilon=0$) and find the time evolution of the inverse kink width, $W(t)$. One example of such a curve is presented in Fig.~\ref{fig3}(a) for $V_k=0.6$, $A=0.25$. It can be seen that $W$ oscillates almost periodically near the value $\overline{W}_0$ with the amplitude $\overline{a}$ and period $\overline{T}$ (or frequency $\overline{\omega}=2\pi/\overline{T}$). These quantities were calculated as the averaged values within the time domain $100<t<200$. We also calculate the actual kink velocity $\overline{V}_k$, which is not equal exactly to the set value $V_k$ due to approximate nature of the ansatz Eq.~(\ref{Ansatz}); the latter is less adequate the larger the excitation of the IM.
The numerically found parameters are compared with the
theoretically predicted parameters given by Eqs.~(\ref{Kinkwidth})-(\ref{Kinkomega}).

In Fig.~\ref{fig3}(b), symbols show $\overline{W}_0$ (triangles) and $\overline{\omega}$ (circles) as functions of $\overline{V}_k$. The solid lines give the predicted dependencies, namely Eq.~(\ref{Kinkwidth}) and Eq.~(\ref{Kinkomega}), respectively. A good coincidence between numerical and predicted values can be seen. Thus, the ansatz of Eq.~(\ref{Ansatz}), even without the $A^2$ term, predicts with a good accuracy the dependence of the IM frequency and width on the
kink velocity.

Panel (c) of Fig.~\ref{fig3} shows the relative difference between the set kink velocity $V_k$ and the actual kink velocity $\overline{V}_k$ as a
function of $V_k$. For small $V_k$ the difference is about 4\%, while for larger velocities it increases up to 7\% within the considered range of $V_k$.

In Fig.~\ref{fig3}(d) the relative difference between numerically found and predicted by Eq.~(\ref{Kinkampl}) values of the kink's IM oscillation amplitude is shown as a function of the set kink velocity $V_k$. For $V_k\leqslant 0.6$ the difference is within 5\% and it grows rapidly for larger kink velocities.

We note that the above described deviation of the actual kink's IM
parameters from the set values is not due to the inaccuracy of the numerical integration of the equations Eq.~(\ref{phi4discrete}), but due to the
inaccuracy of the ansatz Eq.~(\ref{Ansatz}) where, as it was mentioned, the $A^2$ term was not taken into account.

Overall, we conclude that the ansatz of Eq.~(\ref{Ansatz}), even without the
quadratic correction, produces good initial conditions for the kink's IM in a
wide range of kink velocities (except for the large speed relativistic limit).

\subsection {Kink interacting with periodic $\mathcal{PT}$-symmetric perturbation} \label{Sec:IIIB}

The full model of Eqs.~(\ref{phi4})--(\ref{Phi4_Perturbation}) is considered for the perturbation amplitude $\epsilon=0.08$ and different values of the parameter $\beta$. Initially we do not excite the kink's IM and use the ansatz of Eq.~(\ref{Kink}) to boost the kink with the velocity $V_k$. Then the time evolution of the inverse kink width, $W(t)$, and kink velocity, $V_k(t)$, are analyzed in
the course of the numerical integration.

\begin{figure}
\includegraphics[width=9.5cm]{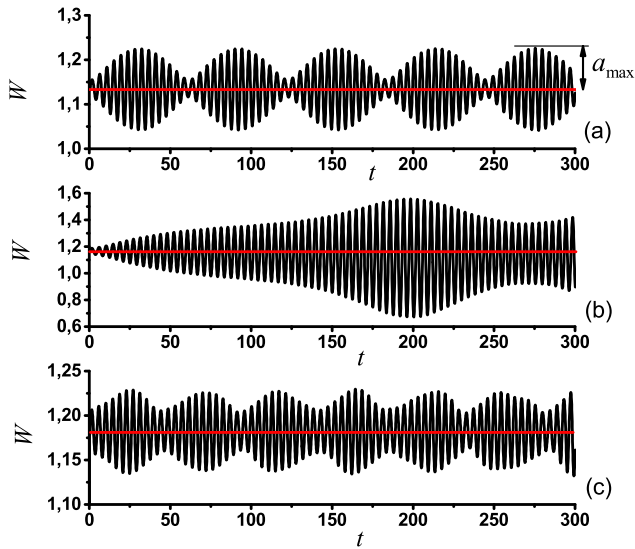}
\caption{Time evolution of the inverse kink width $W(t)$ for $\beta=3$ and initial kink velocities (a) $V_k=0.47$, (b) $V_k=0.51$, and (c) $V_k=0.53$.
  The kink's IM was not excited at $t=0$. Horizontal red lines show the corresponding values of (average inverse kink width) $W_0$.} \label{fig4}
\end{figure}

\begin{figure}
\includegraphics[width=9.5cm]{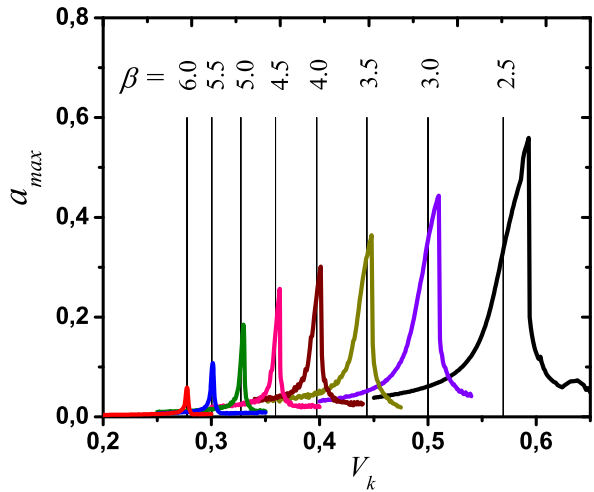}
\caption{Dependence of the maximal IM amplitude as the function of kink velocity for different values of $\beta$ from $2.5$ to $6$ with the step of $0.5$. The perturbation amplitude is $\epsilon=0.08$. The vertical lines show the resonant velocities predicted from Eq.~(\ref{Resonance2}).} \label{fig5}
\end{figure}

Typical examples of the curves for $W(t)$ are given in Fig.~\ref{fig4} for $\beta=3$ and initial kink velocities (a) $V_k=0.47$, (b) $V_k=0.51$, and (c) $V_k=0.53$. Horizontal red lines show the corresponding values of $W_0$ calculated from Eq.~(\ref{Kinkwidth}). It can be seen that the kink inverse width oscillates in time around $W_0$ and the amplitude of the oscillation, which is the kink's IM amplitude, changes in time quasi-periodically. The maximal value of the amplitude, $a_{\max}$, is found within $0\leqslant t \leqslant 400$. The value of $a_{\max}$ strongly depends on the kink velocity. The dependencies of $a_{\max}$ on the kink velocity are presented in Fig.~\ref{fig5} for $\beta$ varying from 2.5 to 6.0 with the step of 0.5. Vertical lines show the resonance values of $V_k$ found for each value of $\beta$ from Eq.~(\ref{Resonance2}). A sharp increase of $a_{\max}$ is found to take place near the predicted values of the resonant velocities.
This clearly suggests that the kink's IM resonantly draws a maximal
amount of energy from the gain-loss environment on which it moves
when its frequency coincides with that of its periodic motion
through a cell of the periodic domain.

\begin{figure}
\includegraphics[width=9.5cm]{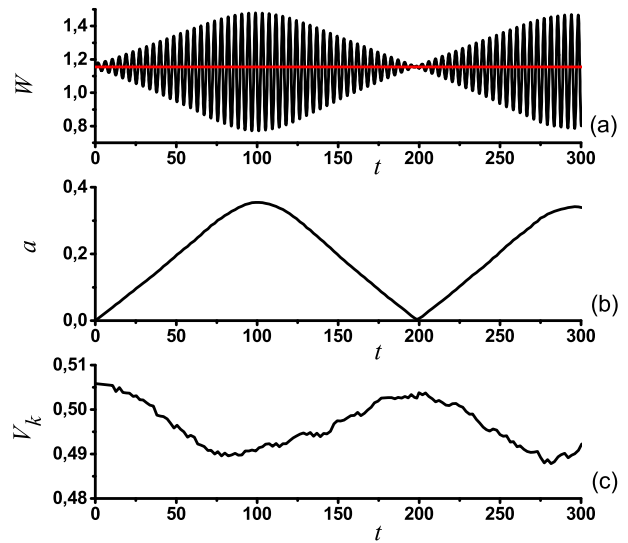}
\caption{Time evolution of (a) the inverse kink width, (b) the IM amplitude [envelope of the function in (a)], and (c) the kink velocity, for $\beta=3$ and initial kink velocity $V_k=0.5$. The horizontal red line in (a) shows the value of $W_0$.} \label{fig6}
\end{figure}

Our next step is to check if there is any energy exchange between the kink translational motion and the kink's IM. In Fig.~\ref{fig6} we plot the time evolution of (a) the inverse kink width, (b) the IM amplitude, and (c) the kink velocity, for $\beta=3$ and initial kink velocity $V_k=0.5$. The horizontal red line in (a) shows the value of $W_0$. Remarkably, when the IM amplitude increases the
kink velocity decreases and thus, the kink translational and vibrational modes are coupled. Of course, this is also expected from the fundamental underlying
premise of~\cite{C3,C4,C5,pgkin,R5}, since it is this exchange of
energy between translational and vibrational which is responsible
for the excitation of the IM in kink collision events (and ultimately
its resonance with the continuous spectrum modes).


\subsection {Effect of the perturbation amplitude $\epsilon$} \label{Sec:IIIC}

\begin{figure}
\includegraphics[width=9.0cm]{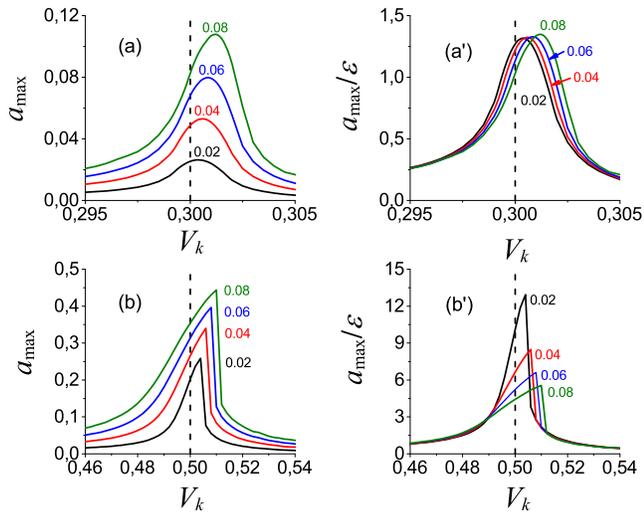}
\caption{Maximal kink's IM amplitude as a function of kink initial velocity for (a) $\beta=5.5$ and (b) $\beta=3$ and for different values of the perturbation parameter $\epsilon$ (indicated for each curve). In (a') and (b') $a_{\max}$ shown in (a) and (b), respectively, is normalized to $\epsilon$. Predicted by Eq.~(\ref{Resonance2}) values of the resonant velocities are shown by the vertical dashed lines.} \label{fig7}
\end{figure}

We calculate maximal kink's IM amplitude as a function of $V_k$ for various values of the perturbation parameter $\epsilon$. The simulations are done for $\beta=5.5$, for which the resonance kink velocity is expected at $V_k=0.3$ [see Eq.~(\ref{Resonance2})] and for $\beta=3$ with the expected resonance peak at $V_k=0.5$. The results are shown in Fig.~\ref{fig7}, where the values of $\epsilon$ are indicated for each curve. In (a), (a') the case of $\beta=5.5$ is shown, while in (b), (b') the results for $\beta=3$ are presented. In (a) and (b) $a_{\max}$ is shown, but in (a') and (b') we present the normalized quantity $a_{\max}/\epsilon$. It can be seen that for relatively small $a_{\max}$ (roughly, for $a_{\max}<0.2$) the effect of perturbation is proportional to $\epsilon$. This is so because the normalized curves in (a') almost merge and so do the normalized curves in (b') in the off-resonance regions. On the other hand, close to the resonance in the case of $\beta=3$, the maximal amplitude is $a_{\max}>0.2$ and the effect of $\epsilon$ becomes nonlinear, as the value of $a_{\max}$ first increases with $\epsilon$ linearly and then demonstrates saturation.

A general trend, which can be realized from Fig.~\ref{fig7}, is that a decrease in the perturbation parameter $\epsilon$ results in the approaching of the resonant peaks to the values predicted from Eq.~(\ref{Resonance2}) and shown by the vertical dashed lines.

\section {Collective Coordinate Method} \label{Sec:IV}

In this section, to gain further insights on the full
numerical results, we will compare those obtained for the continuum
system of Eq.~(\ref{phi4}) with the
collective coordinate method developed in~\cite{KevrekidisRevA}. The $\phi^4$ kink is
effectively described as a two degree of freedom (dof) particle: the two
dof's are the kink coordinate $X(t)$, which takes into account the kink's translational mode, and the kink inverse width $a(t)$. These can be extracted from the following equations~\cite{KevrekidisRevA}
\begin{eqnarray}\label{Coll_var1a}
    M\ddot{X}=\int_{-\infty}^{\infty} (\phi^{\prime}_K
    +aS^{\prime})[(\phi^{\prime}_K
    +aS^{\prime})\dot{X}-\dot{a}S]\gamma(x)dx,
\end{eqnarray}
\begin{eqnarray}\label{Coll_var1aa}
    \ddot{a}=-\omega^2a+\int_{-\infty}^{\infty}
    S[-(\phi^{\prime}_K
    +aS^{\prime})\dot{X}+\dot{a}S]\gamma(x)dx.
\end{eqnarray}
Here $S(x - X_0) =\sqrt{3/2} \tanh(x - X_0){\rm sech}(x -X_0)$ is the kink's
shape mode. The first equation is related to the kink's translational mode and
the latter is associated with the IM  of the $\phi^4$ kink.
These equations yield the general form of the nonconservative
forcing including the coupling between the modes.
Notice, however, that these equations have been derived in~\cite{KevrekidisRevA}
for the case of sufficiently low (i.e., non-relativistic) speeds.
Below we present
the results of numerical solution for the two
degree of freedom model of Eqs.~(\ref{Coll_var1a})-(\ref{Coll_var1aa}).
\begin{figure}
\includegraphics[width=8.0cm]{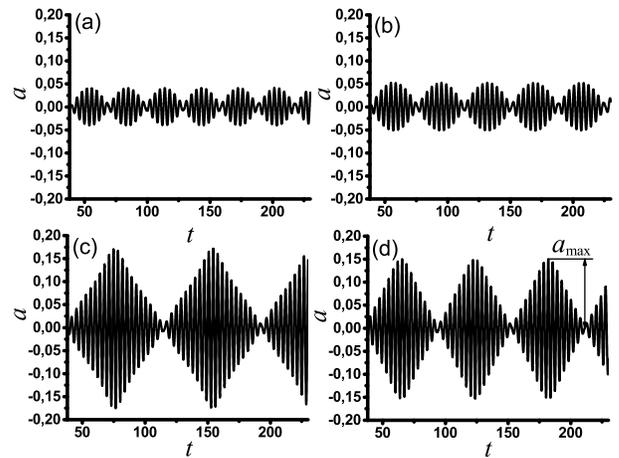}
\caption{Results for the collective variable method of Eqs.~(\ref{Coll_var1a}) and (\ref{Coll_var1aa}). Time evolution of the kink inverse width as a result of the kink interaction with the periodic $\mathcal{PT}$-symmetric perturbation. The kink initial velocity is (a) $V_k=0.47$ (b) $V_k=0.48$ (c) $V_k=0.5$ and (d) $V_k=0.52$. The perturbation parameters are $\epsilon=0.08$ and $\beta=3$.} \label{fig8}
\end{figure}

In Fig.~\ref{fig8}, we plot the kink inverse width as a function of time.
The kink moves with the initial velocity (a) $V_k=0.47$, (b) $V_k=0.48$, (c) $V_k=0.5$ and (d) $V_k=0.52$ interacting with the perturbation characterized by the parameters $\epsilon=0.08$ and $\beta=3$. According to Eq.~(\ref{Resonance2}), for this choice of $\beta$, the resonant kink velocity is equal to $1/2$.
Indeed, for kink velocities near the resonant velocity a noticeable increase of the maximal IM amplitude can be seen. This is in excellent qualitative agreement with the results obtained for the continuum model.

\begin{figure}
\includegraphics[width=9.0cm]{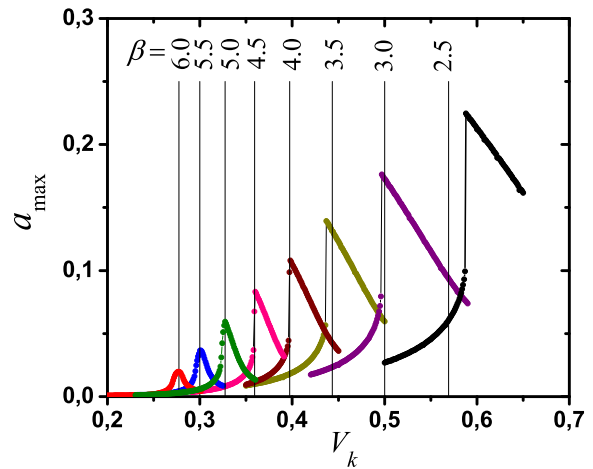}
\caption{ Collective variable results for Eqs.~(\ref{Coll_var1a}) and (\ref{Coll_var1aa}). Maximal kink's IM amplitude as the function of kink initial velocity for different values of $\beta$. The perturbation amplitude is $\epsilon=0.08$. Vertical lines show the resonant velocities predicted from Eq.~(\ref{Resonance2}).} \label{fig9}
\end{figure}

In Fig.~\ref{fig9} we plot the maximal amplitude of the kink's IM as a function of the kink initial velocity for different values of $\beta$ from 2.5 to 6.0 with the step of 0.5. The results are obtained with the collective variable method of Eqs.~(\ref{Coll_var1a}) and (\ref{Coll_var1aa}). The vertical lines show the resonance velocities found from Eq.~(\ref{Resonance2}) for the corresponding values of $\beta$. It can be seen that for a kink velocity close to the resonance velocity the maximal kink's IM amplitude sharply increases. These results are in very good qualitative agreement with numerical simulations of the continuum model (see Fig.~\ref{fig5}). At the quantitative level, however, the collective
coordinate method underestimates the values of $a_{\max}$ (cf. Fig.~\ref{fig5} and Fig.~\ref{fig9}). This is likely related to the fact that the
collective coordinate model of  Eqs.~(\ref{Coll_var1a}) and (\ref{Coll_var1aa})
does not take into account the relativistic effects, in particular, the velocity dependence via the $\delta_k$ factor in the form of the kink, as well
as of the IM frequency given by Eq.~(\ref{Kinkomega}). It would be
a particularly interesting and relevant task to derive the relativistic
corrections of such equations of motion and to examine their
quantitative features in comparison with the full field-theoretic
model results. However, given the unraveling of the physical nature
of the relevant resonances, and the qualitative capturing of all the
relevant phenomenology by the reduced model, we deem this task to
be outside the scope of the present work.

\begin{figure}
\includegraphics[width=9.5cm]{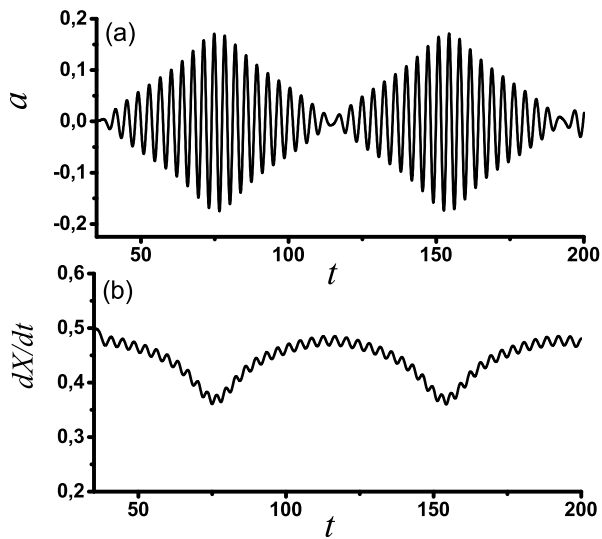}
\caption{ Collective variable results for Eqs.~(\ref{Coll_var1a}) and (\ref{Coll_var1aa}). (a)~Kink's inverse width as the function of time for the kink initial velocity $V=0.5$. (b) Kink velocity as the function of time. The perturbation parameters are $\epsilon=0.08$ and $\beta=3$.} \label{fig10}
\end{figure}

Finally, we examine if the collective variable model predicts the energy exchange between the kink translational and oscillation modes as was found in frame of the continuum model (see Fig.~\ref{fig6}). In Fig.~\ref{fig10} we plot (a) the kink's IM amplitude as a function of time for  $V=0.5$ and (b) the kink velocity as a
function of time.
The perturbation parameters are $\epsilon=0.08$ and $\beta=3$. Very good
qualitative agreement can be seen with the results for the continuum model. The kink velocity decreases when the IM amplitude increases, so that the kink translational and oscillation modes are coupled, as expected and as
discussed previously. In fact, the collective coordinate equations
(\ref{Coll_var1a})-(\ref{Coll_var1aa}) in some sense render this coupling
explicit via their terms involving both $\dot{X}$ (and $\phi_K$) and $a$
(and $S$).

\section {Conclusions} \label{Sec:V}

The interaction of moving $\phi^4$ kinks with periodic $\mathcal{PT}$-symmetric
perturbations was investigated numerically. It has been shown that the kink's IM can be excited with a time-dependent amplitude. The kink inverse width varies in time quasiperiodically, with the maximal amplitude depending on the kink velocity. The maximal IM amplitude sharply increases when the kink velocity is such that it propagates over one period of the gain-loss terrain
in about one period of kink's IM oscillation. This resonant effect is
clearly seen for the kinks having sufficiently large velocity, $V_k>0.25$.
The explanation of this resonance effect is as follows.
When the IM is excited, the kink kinetic energy oscillates in time having two maxima within one period of IM in the case of small kink velocity and only one maximum if the velocity is relatively large. The periodic in space perturbation considered in this work is of the dashpot type with alternating regions of positive and negative damping proportional to $\partial \phi/\partial t$. This means that the effect of the perturbation is maximal when kink's kinetic energy is maximal. Thus, for slowly moving kinks one can expect to observe the resonance effect when the
kink velocity is such that it travels two periods of the perturbative function in one IM period. In this case the period of the perturbative function should be much smaller than the kink width and the effect becomes very weak. For the kinks with relatively large speed, in order to observe the resonance, the period of the perturbed function should be comparable to or even larger than the kink width, and the resonance takes place when the kink travels one period of this function in one IM cycle.

The results of numerical simulations for the continuum perturbed $\phi^4$ model have been compared to the results of the collective coordinate method that takes into account not only the kink's translational but also its
vibrational degrees of freedom~\cite{KevrekidisRevA}. It has been shown that
this theoretical approximation qualitatively predicts all major effects observed for the continuum model.
While the collective coordinate method is not quantitatively accurate,
it is expected that a variant thereof incorporating relativistic
effects is more likely to be. The latter would be an 
interesting direction for future study.

We conclude that the periodic $\mathcal{PT}$-symmetric perturbation gives new
opportunities in the manipulation of the kink dynamics. Particularly, one can control the kink's IM amplitude by varying the parameters of the
perturbation function and the kink velocity. Presumably, such features
are of broader physical interest and relevance than the specifics
of our $\phi^4$ example. In light of the recent realization of
optical solitons in $\mathcal{PT}$-symmetric lattices~\cite{christo},
it would be interesting to extend relevant notions to
nonlinear Schr{\"o}dinger classes of models, especially since
recent work has developed non-conservative variational approximations
in the latter context~\cite{rossi}. These directions are currently
under investigation and will be reported in future publications.

\appendix*
\section{Kink's internal mode}

We consider Eq.~(\ref{phi4}) for $\g=0$. Its kink solution (\ref{Kink}) depends on the special variable $y:=\delta_k(x-x_0-Vt)$. We introduce one more variable as $\tau:=\delta_k(t-V(x-x_0))$. Then we see that by passing from variables $(x,t)$ to $(y,\tau)$ we have
\begin{equation}\label{d2}
\phi_{tt}-\phi_{xx}=\phi_{\tau\tau}-\phi_{yy}
\end{equation}
(i.e., this is the Lorentz invariance)
and this change of variables thus preserves the first two terms in equation (\ref{phi4}) with $\g=0$. Then, it can be rewritten as
\begin{equation}\label{d3}
\phi_{\tau\tau}-\phi_{yy}-2\phi(1-\phi^2)=0,
\end{equation}
while the kink solution becomes $\phi_K(y)=\tanh y$. At the next step we want to construct an approximate solution to (\ref{d3}) in the form
\begin{equation}\label{d5}
\phi_A(y,\tau)=\pm\phi_K(y)+A\phi_1(y,\tau)\pm A^2\phi_2(y,\tau)+\ldots,
\end{equation}
where $A$ is assumed to be a small real parameter, and $\phi_1$, $\phi_2$ are some functions to be determined. By approximate solution we mean the following:

1. The expansion (\ref{d5}) is an asymptotic one, i.e., each next term is smaller by order the previous. In particular, it means that functions $\phi_1$, $\phi_2$ are to be bounded.

2. By substituting the ansatz  (\ref{d5}) into (\ref{d3}) we should get an error of order $O(A^3)$ for small $A$.

The latter condition leads us to the equations for $\phi_1$, $\phi_2$. Namely, substituting (\ref{d5}) into (\ref{d3}) and writing then a power
series expansion w.r.t. $A$, we obtain:
\begin{align*}
&A\big(\phi_{1\tau\tau}-\phi_{1yy}-2(1-3\phi_K^2)\phi_1\big)
\\
&+A^2\big(\phi_{2\tau\tau}-\phi_{2yy}-2(1-3\phi_K^2)\phi_2-6\phi_K\phi_1^2
\big)=O(A^3),
\end{align*}
and therefore,
\begin{align}\label{d6}
& \phi_{1\tau\tau}-\phi_{1yy}-2(1-3\phi_K^2)\phi_1=0,\\
 & \phi_{2\tau\tau}-\phi_{2yy}-2(1-3\phi_K^2)\phi_2=6\phi_K\phi_1^2.
   \label{d7}
\end{align}
We seek special solutions to the above equations by separation of variables:
\begin{equation*}
\phi_1(y,\tau)=e^{\iu\om\tau}\Phi_1(y),\quad \phi_2(y,\tau)=e^{2\iu\om\tau}\Phi_2(y),
\end{equation*}
where $\om$ is some constant. Then equation  (\ref{d6}) becomes
\begin{equation*}
\left(-\frac{d^2\ }{dy^2}+6\tanh^2 y\right)\Phi_1=(\om^2+2)\Phi_1
\end{equation*}
and since $\tanh^2 y=1-\cosh^{-2}y$, we finally have
\begin{equation}\label{d8}
\left(-\frac{d^2\ }{dy^2}-\frac{6}{\cosh^2 y}\right)\Phi_1=(\om^2-4)\Phi_1.
\end{equation}
Equation (\ref{d7}) yields
\begin{equation}\label{d9}
\left(-\frac{d^2\ }{dy^2}-\frac{6}{\cosh^2 y}\right)\Phi_2=(4\om^2-4)\Phi_2+6\Phi_1^2\phi_K(y).
\end{equation}

Equation (\ref{d8}) is the eigenvalue equation for the operator
\begin{equation*}
H:= -\frac{d^2\ }{dy^2}-\frac{6}{\cosh^2 y} \quad\text{in}\quad L_2(\mathds{R}).
\end{equation*}
Its essential spectrum is $[0,+\infty)$ and the equation
\begin{equation*}
\left(-\frac{d^2\ }{dy^2}-\frac{6}{\cosh^2 y}\right)\Phi=0\quad\text{in}\quad\mathds{R}
\end{equation*}
has the only bounded solution
\begin{equation*}
\Phi(y)=\frac{\cosh(2y)-2}{\cosh(2y)+1}.
\end{equation*}
This solution has two zeroes and it means that operator $H$ has two eigenvalues below the essential spectrum. By straightforward calculations one can check that these eigenvalues are $-1$ and $-4$, namely,
\begin{align*}
& \left(-\frac{d^2\ }{dy^2}-\frac{6}{\cosh^2 y}\right)\Psi_1=-4\Psi_1,\quad \Psi_1:=\frac{1}{\cosh^2 y},
\\
& \left(-\frac{d^2\ }{dy^2}-\frac{6}{\cosh^2 y}\right)\Psi_2=-\Psi_2,\quad\hphantom{4} \Psi_2:=\frac{\sinh y}{\cosh^2 y}.
\end{align*}

If we choose $\Phi_1=\Psi_1$ as a solution to (\ref{d7}), it leads us to $w=0$ and this is not the solution with the IM. If we choose $\om\geqslant 4$, then equation (\ref{d8}) has a bounded solution. However, as one can check, then equation (\ref{d9}) has no bounded solution. This is why we choose $\Phi_1=\Psi_2$ and constant $\om$  is non-zero: $\om=\sqrt{3}$. Then for $\Phi_2$ we obtain the equation
\begin{equation}\label{d10}
\left(-\frac{d^2\ }{dy^2}-\frac{6}{\cosh^2 y}-8\right)\Phi_2=  \frac{6\sinh^2 y}{\cosh^5 y}.
\end{equation}
Unfortunately, we can not solve it in elementary functions. Nevertheless, this is an inhomogeneous linear second order ordinary differential equations and its solution can be written by the method of variation of coefficients in terms of the fundamental system of solutions. Thanks to the well-known WKB estimates, these solutions behave at infinity as $e^{\pm\iu\sqrt{8}y}$ and it implies that equation (\ref{d10}) has a bounded solution.

The final form of ansatz (\ref{d5}) expressed in variables $(x,t)$ is as follows:
\begin{equation}
\begin{aligned}
&\phi_A(x,\tau)=\pm \tanh \delta_k(x-x_0-V_kt)
  \\
 &+ Ae^{\iu \sqrt{3} \delta_k  [t-V_k(x-x_0)]} \frac{\sinh\delta_k(x-x_0-V_kt)}{\cosh^2 \delta_k(x-x_0-V_kt)}
  \\
 &\pm A^2e^{2\iu \sqrt{3} \delta_k [t-V_k(x-x_0)]}\Phi_2[\delta_k(x-x_0-V_kt)]
 \\
& +\ldots,
\end{aligned}\label{final}
\end{equation}
where $\Phi_2$ is a solution to equation (\ref{d10}).
As an approximate solution to Eq.~(\ref{phi4}), one should take either real or imaginary part of Eq.(\ref{final}).

\section*{Acknowledgments}
D.S. would like to express special thanks to Prof.~ Kurosh~Javidan for his useful
discussion in this work. D.I.B. was partially supported by a grant of
Russian Foundation for Basic Research (15-31-20037-mol\_ved\_a).
P.G.K. gratefully acknowledges the support of NSF-DMS- 1312856, and the ERC under
FP7, Marie Curie Actions, People, International Research Staff
Exchange Scheme (IRSES-605096). D.S.V. thanks Russian Science Foundation for the financial support under the grant N 16-12-10175.

\end{document}